\documentclass[12pt]{article}

\usepackage{amsmath2000}
\allowdisplaybreaks

\begin{document}

\baselineskip=16.8pt plus 0.2pt minus 0.1pt

\makeatletter
\@addtoreset{equation}{section}
\renewcommand{\theequation}{\thesection.\arabic{equation}}
\renewcommand{\thefootnote}{\fnsymbol{footnote}}

\newcommand{\ap}{\alpha'}
\newcommand{\p}{\partial}
\newcommand{\nn}{\nonumber}
\newcommand{\K}{K_1}
\newcommand{\go}{g_{\rm o}}
\newcommand{\calS}{{\cal S}}
\newcommand{\calQ}{{\cal Q}}
\newcommand{\calEc}{{\cal E}_{\rm c}}
\newcommand{\calNt}{{\cal N}_{\rm t}}
\newcommand{\calQB}{{\cal Q}_{\rm B}}
\newcommand{\calM}{{\cal M}}
\newcommand{\calO}{{\cal O}}
\newcommand{\mt}{m_{\rm t}}
\newcommand{\Phit}{\Phi_{\rm t}}
\newcommand{\vac}{\ket{0}}
\newcommand{\Psic}{\Psi_{\rm c}}
\newcommand{\calNc}{{\cal N}_{\rm c}}
\newcommand{\nt}{n_{\rm t}}
\newcommand{\ds}{\displaystyle}
\newcommand{\Pmatrix}[1]{\begin{pmatrix} #1 \end{pmatrix}}
\newcommand{\sPmatrix}[1]{
            \left(\begin{smallmatrix} #1 \end{smallmatrix}\right)}
\newcommand{\wt}[1]{\widetilde{#1}}
\newcommand{\bm}[1]{\boldsymbol{#1}}
\newcommand{\diag}{\mathop{\rm diag}}
\newcommand{\bra}[1]{\langle #1\vert}
\newcommand{\ket}[1]{\vert #1\rangle}
\newcommand{\braket}[2]{\langle #1\vert #2\rangle}
\newcommand{\tr}{\mathop{\rm tr}}
\newcommand{\Dp}{D_+}
\newcommand{\Dm}{D_-}
\newcommand{\invK}{\frac{1}{\sqrt{K_1^2}}}
\newcommand{\T}{{\rm T}}
\newcommand{\dT}{\Delta T}
\newcommand{\invdT}{\frac{1}{\dT}}
\newcommand{\dU}{\Delta U}
\newcommand{\bmt}{\bm{t}}
\newcommand{\Nt}{{\cal N}_{\rm t}}
\newcommand{\Nc}{{\cal N}_{\rm c}}
\newcommand{\hV}{\widehat{V}}
\newcommand{\wh}[1]{\widehat{#1}}
\newcommand{\calR}{{\cal R}}
\newcommand{\bmu}{\bm{u}}
\newcommand{\bmv}{\bm{v}}
\newcommand{\bz}{b_0}
\newcommand{\QT}{{\cal Q}}
\newcommand{\Ec}{{\cal E}_{\rm c}}
\newcommand{\cvec}[2]{\binom{#1}{#2}}
\newcommand{\D}{D}
\newcommand{\invS}{W}
\newcommand{\calA}{{\cal A}}
\newcommand{\calB}{{\cal B}}
\newcommand{\calC}{{\cal C}}
\newcommand{\calD}{{\cal D}}
\newcommand{\calV}{{\cal V}}
\newcommand{\calW}{{\cal W}}
\newcommand{\SV}{{\cal S}}
\newcommand{\hQ}{\widehat{{\calQ}}}
\newcommand{\C}{{\rm C}}
\newcommand{\R}{R}
\newcommand{\mtFsq}{\left(m_{\rm t}^{\rm Fock}\right)^2}
\newcommand{\mtSsq}{\left(m_{\rm t}^{\rm Sliver}\right)^2}
\newcommand{\hx}{\hat{x}}
\newcommand{\hp}{\hat{p}}
\newcommand{\ld}{{(\lambda)}}
\newcommand{\wtstar}{\,\tilde{\star}\,}
\newcommand{\Slv}{{\rm S}}


\begin{titlepage}
\title{
\hfill\parbox{4cm}
{\normalsize KUNS-1792\\CALT-68-2395\\CITUSC/02-023\\
{\tt hep-th/0206208}}\\
\vspace{1cm}
{\bf Reexamining Classical Solution and Tachyon Mode
in Vacuum String Field Theory}
}
\author{
Hiroyuki {\sc Hata}
\thanks{{\tt hata@gauge.scphys.kyoto-u.ac.jp}}\\
{\it Department of Physics, Kyoto University,
Kyoto 606-8502, Japan}\\[5pt]
\quad and \quad\\[5pt]
Sanefumi {\sc Moriyama}
\thanks{{\tt moriyama@theory.caltech.edu}}\\
{\it California Institute of Technology 452-48,
Pasadena, CA91125, USA}\\[15pt]
}
\date{\normalsize June, 2002}
\maketitle
\thispagestyle{empty}

\begin{abstract}
\normalsize

We reexamine the oscillator construction of the D25-brane solution and
the tachyon fluctuation mode of vacuum string field theory given
previously. Both the classical solution and the tachyon mode are found
to violate infinitesimally their determining equations in the level
cut-off regularization. We study the effects of these violations on
physical quantities such as the tachyon mass and the ratio of the
energy density of the solution relative to the D25-brane tension.
We discuss a possible way to resolve the problem of reproducing the
expected value of one for the ratio.

\end{abstract}

\end{titlepage}

\section{Introduction and summary}

Recently there has been considerable progress in the oscillator
formalism analysis
\cite{RSZ2,HatKaw,Kis,HatMor,MooTay,Oku1,spec,Oku2,Oku3,Okuda,Ima,BarMat}
of vacuum string field theory (VSFT)
\cite{RSZ1,RSZ2,RSZ3,VSFT}, which is a candidate string field theory
expanded around the tachyon vacuum of bosonic open string theory.
The action of VSFT reads
\begin{align}
\SV[\Psi]&=-K\left(
\frac12 \Psi\cdot\calQ\,\Psi+\frac13\Psi\cdot(\Psi * \Psi)\right) ,
\label{eq:SV}
\end{align}
with the BRST operator $\calQ$ given by the following purely ghost
one, respecting that there is no open string excitations around the
tachyon vacuum:
\begin{equation}
\calQ=c_0 + \sum_{n\ge 1}f_n\left(c_n + (-1)^n c_n^\dagger\right) .
\label{eq:calQ}
\end{equation}
In the oscillator formulation of VSFT, there appear infinite
dimensional matrices and vectors in the three-string vertex defining
the $*$-product of two string fields.
Explicitly, the matter part of the vertex reads
\cite{GroJev1,GroJev2}
\begin{align}
&\ket{V_{123}}_{\rm matt}=\left(
\prod_{r=1}^3 \int\frac{d^{d}p_r}{(2\pi)^{d}}\right)
(2\pi)^{d}\delta^d\left(p_1+p_2+p_3\right)
\nn\\
&\qquad\times
\exp
\left\{-\frac12\sum_{n,m\ge 1}
V_{nm}^{rs}a_n^{(r)\dagger}a_m^{(s)\dagger}
-\sum_{n\ge 1}V_{n0}^{rs}a_n^{(r)\dagger}a_0^{(s)}
-\frac12 V_{00}\bigl(a_0^{(r)}\bigr)^2
\right\}\ket{p_1}_1\otimes\ket{p_2}_2\otimes\ket{p_3}_3 ,
\label{eq:Vmatt}
\end{align}
with $d=26$ and $a_0^{(r)}=\sqrt{2}\, p_r$
(we adopt the convention $\ap=1$).\footnote{
In this paper, we adopt a different convention concerning the
center-of-momentum $p$ from that used in
\cite{HatKaw,HatMor,HatMorTer}.
The state $\ket{p}$ is the eigenstate of the momentum operator $\hp$
with eigenvalue $p$ with the normalization
$\braket{p}{p'}=(2\pi)^d\delta^d(p+p')$,
and at the same time it is the
Fock vacuum annihilated by $(a_n,b_n,c_n)$ with $n\ge 1$.
}
We define the Neumann matrices $M_\alpha$ and vectors $\bmv_\alpha$
($\alpha=0,\pm$) by
\begin{equation}
M_0 = C V^{rr},\quad
M_\pm = C V^{r,r\pm 1},\quad
(\bmv_0)_n = V_{n0}^{rr},\quad
(\bmv_\pm)_n = V_{n0}^{r,r\pm1} ,
\label{eq:Mv}
\end{equation}
where $C_{nm}=(-1)^n\delta_{n,m}$ is the twist matrix.
It is also convenient to define the following matrix $M_1$ and vector
$\bmv_1$ which are both twist-odd:
\begin{equation}
M_1=M_+ - M_-,\qquad
\bmv_1= \bmv_+ - \bmv_- .
\label{eq:M1v1}
\end{equation}
These matrices and vectors satisfy various linear and non-linear
identities \cite{GroJev1,GroJev2} which are summarized in appendix
\ref{app:Mvidentities}.
In particular, the matrices $M_\alpha$ are all commutative to each
other.
Similar matrices and vectors appear also in the ghost part of the
three string vertex. We denote them by adding tilde to the
corresponding one in the matter part.

In order for VSFT to really make sense, we have to show the following
two:
\begin{itemize}
\item
There is a classical solution $\Psic$ of VSFT describing a D25-brane,
namely, the perturbative open string vacuum with tachyonic mode.
The energy density $\calEc$ of this solution relative to
that of the trivial one $\Psi=0$ must be equal to the D25-brane
tension $T_{25}$, and the fluctuation modes around $\Psic$ must
reproduce the open string spectrum.

\item
VSFT expanded around $\Psi=0$ describes pure closed string theory
despite that the dynamical variable of VSFT is a open string field.
\end{itemize}

Challenges toward the first problem have been done both by the
oscillator method using the Neumann matrices
and the geometric method using the boundary conformal field theory
(BCFT) \cite{RSZ3,BCFT,GRSZ,RasVis1,RasVis2,butterfly,Oka}.
In the oscillator approach, the classical solution $\ket{\Psic}$ is
assumed to be a squeezed state with its matter part given by
$\exp\bigl(-\frac12\sum_{n,m}a_n^\dagger
\left(CT\right)_{nm}a_m^\dagger\bigr)\ket{0}$. Then the equation of
motion
\begin{equation}
\calQ\Psic +\Psic *\Psic=0 ,
\label{eq:EOMintro}
\end{equation}
is reduced to an algebraic
equation for the matrix $T$, which is solved by using the identities
of appendix \ref{app:Mvidentities} \cite{KosPot,RSZ2}.
The solution obtained this way has been identified \cite{RSZ2,Okuda}
as the sliver state constructed in the BCFT approach \cite{sliver}.
Then, our next task is to construct fluctuation modes around
$\Psic$. This is necessary also for the potential height problem,
namely, the problem of showing $\calEc/T_{25}=1$, since the tension
is given in terms of the open string coupling constant (three tachyon
on-shell coupling) $\go$ by
$T_{25}=1/(2\pi^2\go^2\alpha^{\prime 3})$.
In \cite{HatKaw}, the tachyon fluctuation mode $\Phit$ has been
constructed as a momentum-dependent deformation of $\Psic$:
$\ket{\Phit(p)}=\exp\left(
-\sum_{n\ge 1} t_n a_n^\dagger a_0+ ip\cdot\hx\right)\ket{\Psic}$
which is parameterized by a vector $\bmt$.
This vector $\bmt$ as well as the tachyon mass $\mt^2$ is determined by
the wave equation
\begin{equation}
\calQB\Phit=0 ,
\label{eq:calQBPhit=0intro}
\end{equation}
with $\calQB$ being the BRST
operator around $\Psic$. Owing again to the identities among the
matrices, the equation for $\bmt$ can explicitly be solved.

Once $\Psic$ and $\Phit$ have been found, we can answer the question
of whether the tachyon mass is the expected one, $\mt^2=-1$, and
whether we have $\calEc/T_{25}=1$.
These two physical quantities (we call them observables) are expressed
in terms of the matrices $M_\alpha$ and the vectors $\bmv_\alpha$.
In \cite{HatMor}, a crucial finding has been made concerning these
observables. Both $\mt^2$ and $\calEc/T_{25}$ are given in terms of
quantities (denoted $G$ and $H$ in \cite{HatKaw}) which vanish if we
naively use the non-linear identities of appendix
\ref{app:Mvidentities}, implying absurd results $\mt^2=-\infty$ and
$\calEc/T_{25}=\exp(0/0)/0$.
The vanishing of these quantities $G$ and $H$ can be ascribed to the
fact that the eigenvalues of the matrix $M_0$ are doubly degenerate
between twist-even and odd eigenvectors, and the cancellation occurs
between the contributions of degenerate eigenvalues.
However, since this degeneracy is violated at the end of the
eigenvalue distribution $M_0=-1/3$, and in addition since $G$ and $H$
are singular at $M_0=-1/3$, a careful treatment by using the level
number cut-off regularization leads to finite and non-vanishing values
of $G$ and $H$. This phenomenon that a quantity vanishing naively due
to twist symmetry (eigenvalue degeneracy) can in fact gain a
non-vanishing value has been called ``twist anomaly''
in \cite{HatMor}.

Another important progress concerning the observables in VSFT is that
the eigenvalue problem of the matrices $M_\alpha$ has been solved
in \cite{spec}. They found that the matrices $M_0$ and $M_1$ are
expressed in terms of a simpler matrix $\K$,
\begin{equation}
\bigl(\K\bigr)_{nm}
=-\sqrt{(n-1)n}\,\delta_{n-1,m}-\sqrt{n(n+1)}\,\delta_{n+1,m} ,
\label{K1nm}
\end{equation}
which is the matrix representation of
the Virasoro algebra $\K=L_1+L_{-1}$:
\begin{align}
M_0&=-\frac{1}{1+2\cosh(\K\pi/2)},
\label{eq:M0byK}
\\
M_1&=\frac{2\sinh(\K\pi/2)}{1+2\cosh(\K\pi/2)}.
\label{eq:M1byK}
\end{align}
The matrix $\K$ is symmetric and twist-odd:
$\K^T =\K$ and $C\K C=-\K$.
The eigenvalue distribution of $\K$ is uniform extending from
$-\infty$ to $\infty$. The eigenvalues of $\K$ in the level
cut-off regularization have also been found.

The values of the observables in VSFT were calculated first
numerically \cite{HatKaw,HatMor}, and later analytically
\cite{HatMorTer} by using the results of \cite{spec}.
They were also calculated using the BCFT method in \cite{BCFT}.
The result is that $\mt^2$ is equal to the
expected value of $-1$, but we have a strange value for the ratio;
$\calEc/T_{25}=(\pi^2/3)[16/(27\ln 2)]^3\simeq 2.0558$.
Concerning this problem, a critical observation has been made in
\cite{BCFT}.
In the analysis of \cite{HatKaw,HatMor,HatMorTer}, the wave equation
for the tachyon mode $\Phit$ are implicitly considered in the Fock
space of first-quantized string states, namely, the inner product of
the wave equation with any Fock space elements of the form
$\prod a^\dagger b^\dagger c^\dagger\ket{p}$ are demanded to vanish to
give $\mt^2=-1$.
However, the inner product $\Phit(p)\cdot\calQB\Phit(p)$, which is a
quantity constituting the kinetic term of the tachyon field, no longer
vanishes at $p^2=-\mt^2=1$.
They claimed that this is the reason why we get a wrong value for the
ratio $\calEc/T_{25}$.
Later, the expected value of the ratio, $\calEc/T_{25}=1$, was
successfully derived in \cite{Oka} by introducing a non-linear
component expansion of the string field.
It is interesting to clarify whether the ratio problem can still be
resolved by using the conventional linear expansion.

The purpose of this paper is to reexamine the construction of the
classical solution $\Psic$ and the tachyon wave function $\Phit$
by considering their equations (\ref{eq:EOMintro}) and
(\ref{eq:calQBPhit=0intro}) in the sliver space, namely, the space of
the states of the form
$\ket{\mbox{sliver}}=\prod a^\dagger b^\dagger c^\dagger\ket{\Psic}$.
This is necessary since the tachyon and vector fluctuation modes
around $\Psic$ are in the sliver space \cite{HatKaw,BCFT}.
First, we shall consider how precisely our classical solution $\Psic$
and the tachyon wave function $\Phit$ satisfy (\ref{eq:EOMintro}) and
(\ref{eq:calQBPhit=0intro}), respectively.
In the analysis of twist anomaly, it has been important to refrain
from freely commuting the matrices and using non-linear
identities. Such manipulations lead to wrong results.
However, in solving (\ref{eq:EOMintro}) and
(\ref{eq:calQBPhit=0intro}) for $\Psic$ and $\Phit$, we had to carry
out these potentially invalid operations. We find that both
(\ref{eq:EOMintro}) and (\ref{eq:calQBPhit=0intro}) are
``infinitesimally violated'' by the present $\Psic$ and $\Phit$ in the
level cut-off regularization.

Then, we study the effects of the violations.
The violation of the equation of motion (\ref{eq:EOMintro}) is
invisible so long as we consider (\ref{eq:EOMintro}) in the Fock
space. However, we meet non-vanishing effects of the violation once we
consider the inner product of (\ref{eq:EOMintro}) with the sliver
space elements of the form
$\ket{\mbox{sliver}}=\prod a^\dagger b^\dagger c^\dagger\ket{\Psic}$.
First, the normalization factor of $\Psic$ determined before by
$\braket{\mbox{Fock}}{\calQ\Psic+\Psic *\Psic}=0$ no longer works for
\begin{equation}
\braket{\mbox{sliver}}{\calQ\Psic+\Psic *\Psic}=0 .
\label{eq:sliverEOM}
\end{equation}
Second, there is no unique normalization factor of $\Psic$ common to
all $\bra{\rm{sliver}}$.
We shall carry out similar analysis also for the tachyon wave equation.
We see how the infinitesimal violation of
(\ref{eq:calQBPhit=0intro}), which is invisible in
$\bra{\mbox{Fock}}\calQB\ket{\Phit}$, gives finite effect on
$\bra{\Phit(p)}\calQB\ket{\Phit(p)}$ and makes it non-vanishing at
the expected tachyon on-shell $p^2=1$; a phenomenon pointed out by
\cite{BCFT} using the BCFT arguments.

We also reexamine the potential height problem by considering both
(\ref{eq:EOMintro}) and (\ref{eq:calQBPhit=0intro}) in the sliver
space.
We present an argument which is rather kinematical and needs no
explicit calculation of twist anomalies. We find a kind of no-go
theorem that the ratio is again an undesirable value
$\calEc/T_{25}=\pi^2/(24(\ln 2)^3)\simeq 1.2348$
even if we consider the equations in the sliver space.

Our findings in this paper do not directly help resolve the problem
of the wrong value of the ratio $\calEc/T_{25}$ in the oscillator
approach \cite{HatKaw,HatMor,HatMorTer}. Rather the problem has become
even larger and more complicated by the present analysis.
However, our observation here that the equation of motion in the
sliver space, eq.\ (\ref{eq:sliverEOM}), cannot be satisfied for all
$\bra{\mbox{sliver}}$ reminds us of an interesting proposal of
\cite{Oka} that the string field $\Psi$ in VSFT needs a kind of
non-linear representation.\footnote{
There is another different point between the expression of fluctuation
modes of ours and that of \cite{Oka}.
As has been shown in \cite{BCFT}, our tachyon wave function
corresponds to a local insertion of a tachyon vertex operator on the
sliver surface states in the BCFT language. On the other hand, the
vertex operator is integrated along the surface in the tachyon wave
function of \cite{Oka}.
}
Namely, the space of $\Phi$ is restricted to the sliver space and
expanded around the classical solution as
\begin{equation}
\ket{\Psi}=\ket{\Psic} + \sum_i \varphi_i \ket{i} ,
\label{eq:expandPsi}
\end{equation}
where the summation is running over the sliver space states and the
component fields $\varphi_i$ are not all independent; some of $\varphi_i$
are expressed non-linearly in terms of other and independent $\varphi$
(the component fields $\varphi_1$, $\varphi_2$, $\ldots$ of the
excitations $\ket{{\it 1}}$, $\ket{{\it 2}}$, $\ldots$ depend on one
another in \cite{Oka}).
If we could choose the normalization factor of $\Psic$ in such a way
that (\ref{eq:sliverEOM}) holds for all $\bra{i}$ corresponding to
independent $\varphi_i$, the problem concerning (\ref{eq:sliverEOM})
would be resolved by adopting the non-linear representation.
This could at the same time resolve the problem of $\calEc/T_{25}$
since the kinetic term of the tachyon field has additional
contributions from terms linear in $\varphi_i$ which are not
independent.
This proposal may be interpreted as the string field $\Phi$ being a
constrained one.
If this is the case, it is interesting to consider its relevance to
the problem of whether VSFT around $\Phi=0$ describes a pure closed
string theory.

The organization of the rest of this paper is as follows.
In sec.\ 2, we examine the infinitesimal violation of the equation of
motion of $\Psic$ and its effects in the sliver space.
In sec.\ 3, we present a similar analysis for the wave equation of the
tachyon mode $\Phit$. In sec.\ 4, the potential height problem
is studied by considering both the equation of motion and the wave
equation in the sliver space.
In appendix A, B and C, we present various formulas and technical
details used in the text.

\section{Reexamining classical solution}

In this section, we shall first summarize the oscillator
construction of the D25-brane classical solution $\Psic$ of VSFT and
then examine whether $\Psic$ can be a solution even if we consider the
inner product of the equation of motion with sliver space states.

\subsection{Oscillator construction of the solution}

The D25-brane solution in VSFT is a translationally and Lorentz
invariant classical solution $\Psic$ to the equation of motion of VSFT:
\begin{equation}
\calQ\Psic + \Psic *\Psic=0 .
\label{eq:EOM}
\end{equation}
It has been claimed that such solution $\Psic$ takes a form of
squeezed state in the Siegel gauge \cite{KosPot,RSZ2,HatKaw}:
\begin{equation}
\ket{\Psic(T)}=\calNc\ket{\Slv(T)} ,
\label{eq:Psic=NcS}
\end{equation}
where $\calNc$ is the normalization factor and the squeezed state
$\ket{\Slv(T)}$ is given by
\begin{equation}
\ket{\Slv(T)}= \bz\,\exp\biggl(
-\frac12\sum_{n,m\ge 1}a^\dagger_n(CT)_{nm}a^\dagger_m
+ \sum_{n,m\ge 1}c^\dagger_n(C\wt{T})_{nm}b^\dagger_m \biggr)\vac .
\label{eq:phic}
\end{equation}
The matrices $T$ and $\wt{T}$ are both twist-even, $CTC=T$ and
$C\wt{T}C=\wt{T}$. For the sake of notational simplicity and since we
are mainly interested in the matter part, we have omitted $\wt{T}$ as
an argument of $\ket{\Psic}$ and $\ket{\Slv}$.
Since the three-string vertex $\ket{V_{123}}$ is also of the squeezed
state form, the star product $\Slv(T)*\Slv(T)$ is again a squeezed
state and is given explicitly by
\begin{equation}
\ket{\Slv(T)*\Slv(T)}=\left[\det(1-T\calM)\right]^{-13}
\det(1-\wt{T}\wt{\calM})
\,\hQ(T)\ket{\Slv(T\star T)} ,
\label{eq:S*S}
\end{equation}
where the new matrix $T\star T$ on the RHS is defined by
\begin{equation}
T\star T=M_0+(M_+,M_-)\left(1-T\calM\right)^{-1}T\Pmatrix{M_-\\M_+},
\label{eq:T*T}
\end{equation}
with
\begin{equation}
\calM=\Pmatrix{M_0&M_+\\M_-&M_0}.
\end{equation}
The quantity $\hQ(T)$ on the RHS of (\ref{eq:S*S}) acting on
$\ket{\Slv(T\star T)}$ is
\begin{equation}
\hQ(T)=c_0 + \sum_{n\ge 1}(-1)^n q_n\,c_n^\dagger ,
\label{eq:hQ}
\end{equation}
with the coefficient vector $\bm{q}$ given by
\begin{equation}
\bm{q}=\wt{\bmv}_0+(\wt{M}_+,\wt{M}_-)
\bigl(1-\wt{T}\wt{\calM}\bigr)^{-1}\wt{T}
\cvec{\wt{\bmv}_+}{\wt{\bmv}_-} .
\label{eq:q}
\end{equation}

Therefore, our task of obtaining the solution $\Psic$ has been
reduced to first solving the matrix equation
\begin{equation}
T=T\star T ,
\label{eq:T=T*T}
\end{equation}
for $T$, and then determining the normalization factor $\calNc$ and
the coefficient $\bm{f}$ in the BRST operator $\calQ$ (\ref{eq:calQ})
in such a way that the equation of motion (\ref{eq:EOM}) holds.
Assuming the commutativity $[T,M_\alpha]=0$ and using the non-linear
relations among $M_\alpha$ given in appendix \ref{app:Mvidentities},
in particular, using the formula
\begin{equation}
\left(1-T\calM\right)^{-1}\Bigr|_{\parbox{12ex}{
\scriptsize using non-linear\\ identities}}
=\left(1-2M_0T+M_0T^2\right)^{-1}
\Pmatrix{1-TM_0&TM_+\\ TM_-&1-TM_0},
\label{eq:(1-TM)^-1com}
\end{equation}
valid only when the non-linear identities are used,
the matrix equation (\ref{eq:T=T*T}) is reduced to \cite{KosPot,RSZ2}
\begin{equation}
(T - 1)\Bigl(M_0\,T^2 -(1 + M_0)T+ M_0\Bigr)=0 .
\label{eq:(T-1)()=0}
\end{equation}
As a solution to (\ref{eq:(T-1)()=0}), the following $T^\C$ with
a finite range $[-1,0]$ of eigenvalues has been taken
\begin{equation}
T^\C=\frac{1}{2M_0}\left(1+M_0-\sqrt{(1-M_0)(1+3M_0)}\right) .
\label{eq:T-}
\end{equation}
The equations and the solution for the ghost part matrix $\wt{T}$ are
quite parallel to those for $T$ \cite{HatKaw}.
Finally, the normalization factor $\calNc$ is determined to be given
by
\begin{equation}
\calNc^{\rm Fock}=-\left[\det(1-T\calM)\right]^{13}
\bigl[\det(1-\wt{T}\wt{\calM})\bigr]^{-1} ,
\label{eq:calNcFock}
\end{equation}
and the coefficient $\bm{f}$ in $\calQB$, which is arbitrary for the
gauge invariance of VSFT alone, is fixed to \cite{HatKaw}
\begin{equation}
\bm{f}=\bigl(1-\wt{T}\bigr)^{-1}\bm{q} .
\label{eq:f}
\end{equation}
The reason of the superscript ``Fock'' in $\calNc^{\rm Fock}$
(\ref{eq:calNcFock}) will become clear in the next subsection.

\subsection{Equation of motion in the sliver space}

One might think that the algebraic construction of the solution
$\Psic$ summarized in the previous subsection is quite perfect.
It is, however, a non-trivial problem in what sense the equation of
motion (\ref{eq:EOM}) holds.
The equation of motion holds in the Fock space, namely, we have
\begin{equation}
\braket{\mbox{Fock}}{\calQ\Psic + \Psic *\Psic}=0 ,
\label{eq:FockEOM}
\end{equation}
for any Fock space states of the form
$\ket{\mbox{Fock}}=\prod a^\dagger b^\dagger c^\dagger\ket{0}$.
However, the inner product of the equation of motion
with the solution itself $\ket{\Slv(T)}$ or with the states of the
form $\prod a^\dagger b^\dagger c^\dagger\ket{\Slv(T)}$ (we call such
space the sliver space hereafter) is a non-trivial quantity.
First, let us consider
\begin{equation}
\Slv(T)\cdot\left(\calQ\Psic +\Psic *\Psic\right)=0 .
\label{eq:SEOM}
\end{equation}
Taking into account that the sliver state on the RHS of (\ref{eq:S*S})
is that associated with the matrices $U\equiv T\star T$ and
$\wt{U}\equiv\wt{T}\wtstar\wt{T}$,\footnote{
The product $\wt{T}\wtstar\wt{T}$ for the ghost part matrix is defined
by (\ref{eq:T*T}) with all the matrices including $M_\alpha$ replaced
by the tilded ones.
}
and forgetting (\ref{eq:T=T*T}) for the moment,
we see that the normalization factor $\calNc$ for (\ref{eq:SEOM}) to
hold must be $\calNc^{\rm Sliver}$ related to  $\calNc^{\rm Fock}$
(\ref{eq:calNcFock}) by
\begin{equation}
\frac{\calNc^{\rm Sliver}}{\calNc^{\rm Fock}}
=\left[\frac{\det(1-TU)}{\det(1-T^2)}\right]^{13}
\frac{\det(1-\wt{T}^2)}{\det(1-\wt{T}\wt{U})} .
\label{eq:ratiocalNc}
\end{equation}

Before examining whether the ratio (\ref{eq:ratiocalNc}) is equal to
one, let us next consider the inner product of the equation of motion
with other sliver space elements.
Defining the state $S_J(T)$ by
\begin{equation}
\ket{\Slv_J(T)}=\exp\left(\sum_{n\ge 1}J_n a_n^\dagger\right)
\ket{\Slv(T)} ,
\qquad (J_n^* = J_n) ,
\label{eq:S_J}
\end{equation}
and choosing $\calNc^{\rm Sliver}$ as $\calNc$, we obtain
\begin{align}
&\Slv_J(T)\cdot\left(\calQ\Psic +\Psic *\Psic\right)\Bigr|_{
\calNc=\calNc^{\rm Sliver}}
\nn\\
&\qquad
\propto
\exp\left(-\frac12\bm{J}^\T CT(1-T^2)^{-1}\bm{J}\right)
-\exp\left(-\frac12\bm{J}^\T CU(1-TU)^{-1}\bm{J}\right) .
\label{eq:S_JEOM}
\end{align}
{}From (\ref{eq:S_JEOM}), the inner product of the equation of motion
with, for example, the sliver space state of the form $a^\dagger
a^\dagger\ket{\Slv(T)}$ is given by
\begin{align}
&u_n w_m\left.\frac{\p^2\Slv_J(T)}{\p J_n \p J_m}\right|_{J=0}
\!\!\cdot\left(\calQ\Psic +\Psic *\Psic\right)\Bigr|_{
\calNc=\calNc^{\rm Sliver}}
\propto
\bm{u}^\T C\left[T(1-T^2)^{-1}- U(1-TU)^{-1}\right]\bm{w} ,
\label{eq:SaaEOM}
\end{align}
where $u_n$ and $w_n$ are arbitrary vectors in the level number space.
Recall that we have chosen $\calNc=\calNc^{\rm Sliver}$ in
(\ref{eq:SaaEOM}).

Of course,
the ratio (\ref{eq:ratiocalNc}) is equal to one and the RHS of
(\ref{eq:SaaEOM}) vanishes if we naively use $U\equiv T\star T=T$ and
$\wt{U}\equiv\wt{T}\wtstar\wt{T}=\wt{T}$, which are the equations
determining $T$ and $\wt{T}$.
However, a careful analysis is necessary for them since the matrix
$T^\C$ (\ref{eq:T-}) has eigenvalue $-1$ at the end of its eigenvalue
distribution $[-1,0]$ \cite{HatMor,MooTay,spec}, and each term in
(\ref{eq:ratiocalNc}) and (\ref{eq:SaaEOM}) is singular there. This
implies that we have to introduce a regularization which lifts the
smallest eigenvalue of $T^\C$ from the dangerous point $-1$.

Here, let us adopt the regularization of cutting-off the size of the
infinite dimensional matrices $M_\alpha$ and the vectors $\bmv_\alpha$
to finite size ($L\times L$) ones. This is the regularization we used
in calculating observables (twist anomaly) of VSFT
\cite{HatKaw,HatMor,HatMorTer}, and the following
analysis for (\ref{eq:ratiocalNc}) and (\ref{eq:SaaEOM}) is quite
similar to those for twist anomaly.
As given in (\ref{eq:M0byK}) and (\ref{eq:M1byK}), $M_0$ and $M_1$ are
expressed in terms of a single infinite dimensional matrix $\K$,
and the dangerous eigenvalue $T^\C=-1$ ($M_0=-1/3$) corresponds to
$\K=0$.
In our regularization, we regard $M_0$ and $M_1$ as primary and
hence cut off (\ref{eq:M0byK}) and (\ref{eq:M1byK}) after evaluating
them by using infinite dimensional $\K$.\footnote{
If we adopt the regularization of replacing $\K$ in
(\ref{eq:M0byK}) and (\ref{eq:M1byK}) by a finite size one $\K|_L$,
$M_0$ and $M_1$ remain commutative to each other and hence all twist
anomalies in VSFT vanish identically.
}
Since the neighborhood of $\K=0$ is important for
(\ref{eq:ratiocalNc}) and (\ref{eq:SaaEOM}), let us expand
$M_\alpha$ in our regularization in powers of $\K$:
\begin{align}
&M_0|_L \simeq -\frac13 + \frac{\pi^2}{36}\K^2|_L ,
\label{eq:expanM0}
\\
&M_1|_L \simeq \frac{\pi}{3}\K|_L ,
\label{eq:expanM1}
\end{align}
where $\K^2|_L$ in (\ref{eq:expanM0}) stands for the cut-off of the
infinite dimensional matrix $(\K)^2$.
Similarly, $T^\C$ (\ref{eq:T-}) in our regularization is
expanded as follows:
\begin{equation}
T^\C \simeq -1 +\sqrt{3}\sqrt{1+ 3M_0|_L}
\simeq -1 +\frac{\pi}{2}\sqrt{\K^2|_L} .
\label{eq:expanT^C}
\end{equation}
By the present regularization, the dangerous eigenvalue $-1$ of $T^\C$
is lifted by an amount of order $1/\ln L$ \cite{spec,HatMorTer}.

In our regularization, the matrices $M_0|_L$ and $M_1|_L$ no longer
satisfy the non-linear identities given in appendix
\ref{app:Mvidentities}.
This implies that the equation (\ref{eq:T=T*T}) determining $T$ is not
exactly satisfied by the regularized $T^\C$.
For studying the effects of this violation of (\ref{eq:T=T*T}),
let us consider the expansion of $T\star T$ in powers of $\K$.
Namely, we substitute (\ref{eq:expanM0}), (\ref{eq:expanM1}) and
\begin{equation}
T\simeq -1 + \dT ,
\label{eq:T=-1+dT}
\end{equation}
with $\dT$ being of order $\K$ into the original definition
(\ref{eq:T*T}) of $T\star T$, and calculate it to order $\K$ by
keeping the ordering of the matrices.
The details of the calculation is given in appendix
\ref{app:derivation}, and the result is
\begin{equation}
U=T\star T\simeq -1 +\frac12\dT +\frac98M_1\invdT M_1
\simeq -1 + \frac12\dT +\frac{\pi^2}{8}\K\invdT\K ,
\label{eq:expanT*T}
\end{equation}
where $\K$ and $\dT$ should be regarded as cut-off ones.
Eq.\ (\ref{eq:expanT*T}) is valid for any $\dT$ of order $\K$, not
restricted to $(\pi/2)\sqrt{\K^2|_L}$ corresponding to
(\ref{eq:expanT^C}).
Eq.\ (\ref{eq:expanT^C}), namely,
\begin{equation}
\dT^\C=\frac{\pi}{2}\sqrt{\K^2|_L} ,
\label{eq:dT^C}
\end{equation}
and (\ref{eq:expanT*T}) implies that the equation determining $T$,
(\ref{eq:T=T*T}), is indeed not satisfied by $T=T^\C$ near $\K=0$
since we have $\K^2|_L\ne\left(\K|_L\right)^2$.

Although the violation of (\ref{eq:T=T*T}) we have found here is
infinitesimal since we are considering the range of infinitesimal
eigenvalues of $\K$ of order $1/\ln L$ \cite{spec,HatMorTer},
it can give finite effects on (\ref{eq:ratiocalNc}) and
(\ref{eq:SaaEOM}).
First, the RHS of (\ref{eq:SaaEOM}) is nothing but of the form of
twist anomaly in VSFT discussed in \cite{HatMor,HatMorTer}.
It vanishes naively and has
degree of singularity three if the vectors $\bm{u}$ and $\bm{w}$ are
twist-odd ones with degree of singularity equal to one such as
$\bmv_1$.\footnote{
See \cite{HatMor,HatMorTer} for the degree of singularity
in the calculation of twist anomalies.
We present it in table \ref{tab:degree} for various quantities.
If the degree of singularity of a quantity is less than three, naive
manipulations using the non-linear identities are allowed. Twist
anomaly is a quantity which has degree of divergence equal to three
and vanishes if we use the non-linear identities.
}
\renewcommand{\arraystretch}{1.3}
\begin{table}[htbp]
\begin{center}
\begin{tabular}[b]{|c|c|c|c|c|}
\hline
$1/\sqrt{1+3M_0},1/\sqrt{\K^2}$&$M_1,\K$&$\bmv_0$&$\bmv_1$&$\bm{t}$\\
\hline\hline
$1$&$-1$&$0$&$1$&$1$\\
\hline
\end{tabular}
\caption{Degrees of singularity for various quantities.}
\label{tab:degree}
\end{center}
\end{table}
\renewcommand{\arraystretch}{1}
Therefore, the value of the RHS of  (\ref{eq:SaaEOM}) is
correctly calculated by substituting the expansions
\begin{equation}
(1-T^2)^{-1}\simeq \frac{1}{2\dT} ,\quad
(1-TU)^{-1}\simeq \frac{1}{2\dT}\R ,
\label{eq:expan(1-TU)^-1}
\end{equation}
with $\R$ defined by
\begin{equation}
\R\equiv\left(\frac34 +\frac{\pi^2}{16}\K\invdT\K\invdT\right)^{-1} ,
\label{eq:R}
\end{equation}
and keeping only the most singular part. We have
\begin{equation}
\bm{u}^\T C\left[T(1-T^2)^{-1}- U(1-TU)^{-1}\right]\bm{w}
=\frac12\bm{u}^\T\invdT\left(1-\R\right)\bm{w} .
\label{eq:evalSaaEOM}
\end{equation}
As we experienced in the analysis of twist anomaly, this is indeed
finite and non-vanishing for generic twist-odd $\bm{u}$ and $\bm{w}$
with degree of singularity equal to one (for example,
$u_n=w_n=\delta_{n,1}$).

As for the ratio (\ref{eq:ratiocalNc}) we do not know whether a
similar treatment of taking only the most singular part, for example,
\begin{equation}
\left.\frac{\det(1-TU)}{\det(1-T^2)}\right|_{\rm most\ singular\ part}
=\det \R^{-1} ,
\label{eq:msp}
\end{equation}
can correctly reproduce the original value.
However, there is no reason to believe that the ratio can
keep the value one even though it is non-trivial near
$\K=0$. Numerical analysis of the ratio also supports a value largely
deviated from one \cite{HatKaw}.

Let us summarize the observations made in this subsection.
The sliver state solution $\Psic(T)$ with $T$ given by $T^\C$
(\ref{eq:T-}) and $\calNc$ by $\calNc^{\rm Fock}$ (\ref{eq:calNcFock})
satisfies the equation of motion in the Fock space, (\ref{eq:FockEOM}).
However, this $\Psic(T)$ does not satisfy the equation of motion in
the sliver space. Even if we choose another $\calNc$,
$\calNc^{\rm Sliver}$ given by (\ref{eq:ratiocalNc}), for which the
inner product of the equation of motion with $\Psic(T)$ itself
vanishes, the inner products with other sliver space states fail to
vanish in general.
Namely, there exists no universal $\calNc$ for which the equation
motion holds in the whole sliver space.

The origin of this trouble concerning the classical solution is
the fact that $T^\C$ infinitesimally violates the basic
equation $T=T\star T$ (\ref{eq:T=T*T}) in our level truncation
regularization. Namely, $\dT^\C$ of (\ref{eq:dT^C}) does not satisfy
the $O(\K)$ part of (\ref{eq:T=T*T}) obtained from (\ref{eq:T=-1+dT})
and (\ref{eq:expanT*T}):
\begin{equation}
\dT=\frac{\pi^2}{4}\K\invdT\K .
\label{eq:idealeq}
\end{equation}
One might think that the troubles we have seen above are resolved once
we find an ideal $\dT$ which does satisfy (\ref{eq:idealeq}).
We shall see in the next section that, even if there is such $\dT$
satisfying (\ref{eq:idealeq}), it leads to completely uninteresting
results concerning the observables in VSFT.
Here we point out that (\ref{eq:idealeq}) cannot completely fix the
matrix $\dT$. To see this,
let us move to the representation of matrices in the level number
space where the odd indices are in the upper/left block and the even
ones in the lower/right block.
Taking into account that $\dT$ is twist-even while $\K$ is twist-odd
(both are symmetric), $\dT$ and $\K$ in this representation are
expressed as follows:
\begin{align}
&\dT=\Pmatrix{\dT_{oo} & 0 \\ 0 & \dT_{ee}} ,
\label{eq:dToe}
\\
&\K=\Pmatrix{0 & (\K)_{oe} \\ (\K)_{eo} & 0},
\quad
(\K)_{eo}=(\K)_{oe}^\T .
\label{eq:Koe}
\end{align}
Substituting these expressions into (\ref{eq:idealeq}), we obtain only
one independent equation relating $\dT_{oo}$ and $\dT_{ee}$:
\begin{equation}
\dT_{oo}=\frac{\pi^2}{4}(\K)_{oe}\left(\dT_{ee}\right)^{-1}
(\K)_{eo} .
\label{eq:dToobydTee}
\end{equation}
In the particular case that the size $L$ of the regularized matrices
is an odd integer, $L=2\ell +1$, the relation (\ref{eq:dToobydTee})
shows that (\ref{eq:idealeq}) is self-contradictory.
In this case, $(\K)_{eo}$ is a rectangular $\ell\times(\ell+1)$ matrix
and necessarily has a zero-mode.
Eq.\ (\ref{eq:dToobydTee}) implies that this zero-mode is at the same
time a zero-mode of $\dT_{oo}$ and hence the inverse $1/\dT$ does not
exist.

\section{Reexamining the tachyon wave function}

In the previous section, we saw how the infinitesimal violation of the
basic equation (\ref{eq:T=T*T}) leads to difficulties of
the equation of motion (\ref{eq:EOM}) in the sliver space.
In this section we shall examine the same kind of violation of the
wave equation for the tachyon fluctuation mode.

\subsection{Tachyon wave function in the Fock space}
The tachyon (in general a physical state) fluctuation mode $\Phit$
should satisfy the wave equation, namely, the linearized equation of
motion of the fluctuation:
\begin{equation}
\calQB\Phit \equiv
\calQ \Phit + \Psic *\Phit + \Phit *\Psic=0 ,
\label{eq:calQBPhit=0}
\end{equation}
where $\calQB$ is the BRST operator around the classical solution
$\Psic$ which we expect to describe a D25-brane.
In this subsection, we shall recapitulate the construction of
$\Phit$ given in \cite{HatKaw}.
Like in the case of the classical solution $\Psic$, it is a
non-trivial matter in which space the wave equation
(\ref{eq:calQBPhit=0}) holds. As we shall see in the next subsection,
the tachyon mode $\Phit$ constructed here satisfies
(\ref{eq:calQBPhit=0}) in the Fock space without any problem:
\begin{equation}
\bra{\mbox{Fock}}\calQB\ket{\Phit}=0 .
\label{eq:FockWE}
\end{equation}
However, there will arise subtle issues if we consider
(\ref{eq:calQBPhit=0}) in the sliver space.

In \cite{HatKaw} the following form has been assumed for the tachyon
mode $\Phit$ carrying the center-of-mass momentum $p_\mu$:
\begin{equation}
\ket{\Phit}=\calNt\ket{\Slv(T,\bmt,p)} ,
\label{eq:Phit=calNtS}
\end{equation}
where $\calNt$ is a normalization factor (which is irrelevant for
(\ref{eq:calQBPhit=0}) alone) and the state
$\ket{\Slv(T,\bmt,p)}$ is a $p_\mu$-dependent deformation of the
sliver state $\ket{\Slv(T)}$:
\begin{equation}
\ket{\Slv(T,\bmt,p)}=\exp\biggl(-\sum_{n\ge 1}t_n a_n^\dagger a_0
+ip\cdot \hat{x}\biggr)\ket{\Slv(T)} .
\label{eq:S(T,t,p)}
\end{equation}
The state $\ket{\Slv(T,\bmt,p)}$ depends on a vector $\bmt=(t_n)$ which
is $C$-even, $C\bmt=\bmt$, since the tachyon state is twist-even.
Then the following formula holds for the $*$-product of the sliver
state and the present tachyon mode:
\begin{align}
\ket{\Slv(T)*\Slv(T,\bmt,p)}&=\frac{\det(1-\wt{T}\wt{\calM})}{
\left[\det(1-T\calM)\right]^{13}}
\exp\left(-\frac12 G(T,\bmt)(a_0)^2\right)
\hQ(T)\ket{\Slv(T\star T,T\star\bmt,p)} ,
\label{eq:S(T)*S(T,t,p)}
\end{align}
where the constant $G(T,\bmt)$ and a new vector $T\star\bmt$ are
defined respectively by
\begin{align}
G(T,\bmt)&=2V_{00}
+(\bmv_--\bmv_+,\bmv_+-\bmv_0)\frac{1}{1-T\calM}\left[
T\Pmatrix{\bmv_+-\bmv_-\\\bmv_--\bmv_0}
+2\Pmatrix{0\\\bmt}\right]
\nn\\
&\qquad\quad
+(0,\bmt)\calM\frac{1}{1-T\calM}\Pmatrix{0\\\bmt} ,
\label{eq:G}
\\
T\star\bmt &=\bmv_0-\bmv_+
+(M_+,M_-)\frac{1}{1-T\calM}
\biggl[T\Pmatrix{\bmv_+-\bmv_-\\\bmv_--\bmv_0}
+\Pmatrix{0\\\bmt}\biggr] .
\label{eq:T*t}
\end{align}
Therefore, the wave equation (\ref{eq:calQBPhit=0}) holds at
\begin{equation}
p^2 = -\mtFsq \equiv \frac{\ln 2}{G(T,\bmt)} ,
\label{eq:mtFsq}
\end{equation}
provided $\calNc$ is given by $\calNc^{\rm Fock}$
(\ref{eq:calNcFock}), the matrix $T$ satisfies (\ref{eq:T=T*T}), and
the vector $\bmt$ is a solution to
\begin{equation}
T\star\bmt =\bmt .
\label{eq:T*t=t}
\end{equation}
In \cite{HatKaw}, by adopting $T=T^\C$ and freely using the non-linear
relations among the matrices, the following vector $\bmt^\C$ was
taken as a solution to (\ref{eq:T*t=t}):
\begin{align}
\bmt^\C &=3(1+T^\C)(1+3 M_0)^{-1}\bmv_0
\nn\\
&=-(1+T^\C)\left[(1-M_0)(1+3M_0)\right]^{-1}M_1\bmv_1
\simeq -\frac32\invK\K\bmv_1 ,
\label{eq:t^C}
\end{align}
In (\ref{eq:t^C}), the second expression is due to
$\bmv_0=-(1/3)(1-M_0)^{-1}M_1\bmv_1$ from (\ref{Mv}), and the last
approximate expression has been obtained by substituting
the expansions (\ref{eq:expanM0}), (\ref{eq:expanM1}) and
(\ref{eq:expanT^C}) and keeping only the leading term with degree of
singularity one.
It has been shown numerically in \cite{HatKaw,HatMor} and analytically
in \cite{HatMorTer} that the expected value of the tachyon mass,
$\mtFsq=-1$ is obtained for $\bmt^\C$. Namely, we have
\begin{equation}
G(T^\C,\bmt^\C)=
\frac{9}{4\pi}\bmv_1^\T\biggl(
\invK-\K\Bigl(\invK\Bigr)^3\K\biggr)\bmv_1
=\ln 2 .
\label{eq:G(T^C,t^C)}
\end{equation}

\subsection{Reexamination of the tachyon mode}

Let us reexamine how precisely the equation determining $\bmt$,
eq.\ (\ref{eq:T*t=t}), is satisfied by
$\bmt^\C$ (\ref{eq:t^C}) in the level cut-off regularization.
Using the formulas in appendix \ref{app:derivation}, in particular,
(\ref{eq:1/(1-TM)base}) and (\ref{eq:expanUpm}), we can show that the
degree of singularity one part of $T\star\bmt$ is given for a generic
$T$ with the expansion (\ref{eq:T=-1+dT}) and a generic $\bmt$ with
degree of singularity one by
\begin{align}
&\frac{1+C}{2}\,T\star\bmt\simeq
-\frac{3\pi}{8}\K\invdT\bmv_1 +\frac12\,\bmt ,
\label{eq:T*teven}
\\
&\frac{1-C}{2}\,T\star\bmt\simeq
-\frac34\bmv_1 -\frac{\pi}{4}\K\invdT\bmt .\
\label{eq:T*todd}
\end{align}
Therefore, the $C$-even part and the odd one of (\ref{eq:T*t=t})
restricted to the degree of singularity one part read respectively
\begin{align}
&\bmt\simeq -\frac{3\pi}{4}\K\invdT\bmv_1 ,
\label{eq:T*t=teven}\\
&\frac{\pi}{4}\K\invdT\bmt + \frac34\bmv_1\simeq 0 .
\label{eq:T*t=todd}
\end{align}
Taking $\dT=\dT^\C$ (\ref{eq:dT^C}), we see that the last expression
of (\ref{eq:t^C})  satisfies neither (\ref{eq:T*t=teven}) nor
(\ref{eq:T*t=todd}), though of course they are satisfied if we are
allowed to carry out naive calculations by forgetting that the
matrices are the regularized ones.

The violation of the degree of singularity one part of
(\ref{eq:T*t=t}) observed above, does not invalidate the fact that the
wave equation in the Fock space, (\ref{eq:FockWE}), is satisfied by
$T^\C$ and $\bmt^\C$ at $p^2=1$.\footnote{
In order for (\ref{eq:FockWE}) to hold for any Fock space state
$\ket{\rm Fock}$, $\bmu\cdot(\bmt-T\star\bmt)=0$ must hold for any
normalizable vector $\bmu$. Since the degree of singularity of such
$\bmu$ is at most one and hence the degree of the whole inner product
is at most two, naive manipulations are allowed for calculating this
inner product to give zero.
In this sense, the violation of (\ref{eq:T*t=t}) found here is
infinitesimal.
}
However, let us consider whether a better choice of $\bmt$ is possible
which fully satisfies both (\ref{eq:T*t=teven}) and
(\ref{eq:T*t=todd}).
The degree one part of such an ideal $\bmt$ should be given by
(\ref{eq:T*t=teven}) for a chosen $\dT$. In order for the
second equation (\ref{eq:T*t=todd}) to be consistent, $\dT$ must
satisfy (\ref{eq:idealeq}), which is the $O(\K)$ part of
(\ref{eq:T=T*T}). Therefore, an ideal $\bmt$ exists for an ideal
$\dT$. Unfortunately, we have $G=0$ for such an ideal solution,
implying that $-\mtFsq=\infty$.
This is seen as follows.
Keeping in $G$ (\ref{eq:G}) only those terms with degree of
singularity equal to three by the help of the formulas (\ref{eq:1/(1-TM)base})
and (\ref{eq:expanUpm}), we obtain the following concise expression
of $G$ valid for any $1/\dT$ and $\bmt$ which are both twist-even and
carry degree of singularity one:
\begin{equation}
G(T,\bmt)=
\frac98\,\bmv_1^\T\invdT\bmv_1- \frac12\bmt^\T\invdT\bmt
+G_{\rm reg} ,
\label{eq:simpleG}
\end{equation}
where the last term $G_{\rm reg}$ with degree less three should be
determined from the requirement that the whole of the RHS of
(\ref{eq:simpleG}) vanishes by naive manipulations.
Plugging (\ref{eq:T*t=teven}) into (\ref{eq:simpleG}) and using
(\ref{eq:idealeq}), we get
\begin{equation}
G(T^{\rm ideal},\bmt^{\rm ideal})
=\frac98\bmv_1^\T\invdT\left(
\dT-\frac{\pi^2}{4}\K\invdT\K\right)\invdT\bmv_1=0 .
\label{eq:Gideal=0}
\end{equation}
Therefore, we do not obtain physically sensible results if both the
equations (\ref{eq:T=T*T}) and (\ref{eq:T*t=t})
hold rigorously.\footnote{
If we respect only (\ref{eq:T*t=teven}) and take $T=T^\C$,
the corresponding value of $G$ no longer reproduces
$\mtFsq=-1$. Namely, we have
$$
G\Bigl(T^\C,\bmt\simeq -(3/2)\K\invK\bmv_1\Bigr)=
\frac{9}{4\pi}\,\bmv_1^\T\invK\biggl(
1-\K\invK\K\invK\biggr)\bmv_1
=0.42\ldots
$$
Such $\bmt$ is obtained from the second expression of
$\bmt^\C$ (\ref{eq:t^C}) by moving $M_1$ to the left most position
$$
-M_1(1+T^\C)\left[(1-M_0)(1+3M_0)\right]^{-1}\bmv_1
\simeq -\frac32\K\invK\,\bmv_1 .
$$
Therefore, we could say that the result (\ref{eq:G(T^C,t^C)}) and
hence $\mtFsq=-1$ is owing to an accidental choice of the ordering of
the matrices in $\bmt^\C$ (\ref{eq:t^C}) made in \cite{HatKaw}.
}

Let us return to generic $T$ and $\bmt$ which do not necessarily
satisfy (\ref{eq:T=T*T}) and (\ref{eq:T*t=t}).
In \cite{HatKaw,HatMor,HatMorTer}, the tachyon wave equation is
implicitly considered in the Fock space.
However, recalling that the original string field $\Psi$ in VSFT
is expanded around a classical solution $\Psic$ as
\begin{equation}
\ket{\Psi}=\ket{\Psic} + \sum_i\int\!\frac{d^{26}p}{(2\pi)^{26}}
\,\varphi_i(p)\ket{\Phi_i(p)} ,
\label{eq:Psi=Psic+}
\end{equation}
with $\ket{\Phi_i}$ and $\varphi_i$ being the fluctuation wave
function and the corresponding component field (dynamical variable)
for the open string mode $i$ ($i=$ tachyon, massless vector, etc.), it
is necessary to examine the wave equation in
the sliver space, or more specifically the inner product
$\Phi_t\cdot\calQB\Phi_t$ if we can choose $\Phi_i$ for other modes so
that the mixing with $\Phit$ vanishes, $\Phi_i\cdot\calQB\Phi_t=0$.
In the rest of this section we shall show how the tachyon mass in the
sliver space determined by
\begin{equation}
\Phi_t\cdot\calQB\Phi_t=0 ,
\label{eq:PhitcalQBPhit=0}
\end{equation}
can differ from that in the Fock space, (\ref{eq:mtFsq}).
We shall see that this discrepancy is again due to the infinitesimal
violations of (\ref{eq:T=T*T}) and (\ref{eq:T*t=t}).

To identify the tachyon mass from (\ref{eq:PhitcalQBPhit=0}), let us
calculate its LHS by first calculating $\calQB\Phit$, in particular,
$\Psic *\Phit+\Phit *\Psic$, and then taking its inner product with
$\Phit$.\footnote{
If we calculate directly
$\bra{\Phit}\bra{\Psic}\braket{\Phit}{V}$ like in the evaluation of
the three-tachyon coupling $\go$ \cite{HatKaw}, we obtain the same
result without referring to $T\star T$ nor $T\star\bmt$.
This way of calculation is essentially carried out in sec.\ 4.
The tachyon mass squared (\ref{eq:mtbyVW}) obtained there agrees with
(\ref{eq:mtSsq}) since we have $2\calW-\calV=2 H$
\cite{HatKaw,HatMor,HatMorTer}.
}
Note first the
following equation obtained by taking the inner product between
$\Slv(T,\bmt,p')$ and (\ref{eq:S(T)*S(T,t,p)}):
\begin{align}
\Slv(T,\bmt,p')\cdot\bigl(\Slv(T)*\Slv(T,\bmt,p)\bigr)
&=\frac{\det(1-\wt{T}\wt{\calM})}{\left[\det(1-T\calM)\right]^{13}}
\frac{\det(1-\wt{T}\wt{U})}{\left[\det(1-TU)\right]^{13}}
\nn\\
&\quad\times
\exp\left(-\frac12\left(G+A\right)(a_0)^2\right)
(2\pi)^{26}\delta^{26}(p+p') ,
\label{eq:S.S*S}
\end{align}
with $A(T,\bmt)$ defined by
\begin{equation}
A(T,\bmt)=(T\star\bmt)^\T(1-TU)^{-1}TC(T\star\bmt)
+\bmt^\T U(1-TU)^{-1}\bmt -2\,(T\star\bmt)^\T(1-TU)^{-1}\bmt .
\label{eq:A}
\end{equation}
Using (\ref{eq:S.S*S}) and (\ref{eq:ratiocalNc}), we find that the
LHS of (\ref{eq:PhitcalQBPhit=0}) is given by
\begin{align}
&\Phit(T,\bmt,p')\cdot\calQB\Phit(T,\bmt,p)
=\Phit(T,\bmt,p')\cdot\calQ\Phit(T,\bmt,p)
\left[
1 - 2\frac{\calNc}{\calNc^{\rm Sliver}}
\exp\Bigl(-H(T,\bmt)(a_0)^2\Bigr)
\right] ,
\end{align}
where $H(T,\bmt)$ is
\begin{equation}
H(T,\bmt)=\frac12\left(G + A\right)(T,\bmt)
+\bmt^\T(1+T)^{-1}\bmt
=\frac12 G +\frac12\Bigl(
A - A\bigr|_{(U,T\star\bmt)\to(T,\bmt)}\Bigr) .
\label{eq:H}
\end{equation}
Therefore, if we take $\calNc^{\rm Sliver}$ (\ref{eq:ratiocalNc}) as
the normalization factor $\calNc$ for $\Psic$, the tachyon mass
determined by (\ref{eq:PhitcalQBPhit=0}) is given in terms of $H$ by
\begin{equation}
-\mtSsq \equiv\frac{\ln 2}{2 H(T,\bmt)} .
\label{eq:mtSsq}
\end{equation}

The expression (\ref{eq:H}) tells us that $H$, like $G$, vanishes for
ideal $T$ and $\bmt$ satisfying (\ref{eq:T=T*T}) and (\ref{eq:T*t=t})
without infinitesimal violations, i.e.\  we have
$H(T^{\rm ideal},\bmt^{\rm ideal})=0$.
If the state $\Slv(T\star T,T\star\bmt,p)$ on the RHS of
(\ref{eq:S(T)*S(T,t,p)}) were replaced with $\Slv(T,\bmt,p)$,
the difference $A - A\bigr|_{(U,T\star\bmt)\to(T,\bmt)}$ in
(\ref{eq:H}) would vanish and the Fock space result (\ref{eq:mtFsq})
would be recovered. However, the difference can be non-vanishing since
the infinitesimal discrepancy between
$\bmt$ and $T\star\bmt$ and that between $T$ and $U\equiv T\star T$
are amplified due to the singularity of $(1-TU)^{-1}$ at $T=U=-1$.
Using (\ref{eq:T=-1+dT}), (\ref{eq:expanT*T}),
(\ref{eq:expan(1-TU)^-1}), (\ref{eq:T*teven}) and (\ref{eq:T*todd}),
we obtain
\begin{align}
A - A\bigr|_{(U,T\star\bmt)\to(T,\bmt)}
=&\frac{9}{32}\bmv_1^\T\invdT\left(
1-\frac{\pi^2}{4}\K\invdT\K\invdT\right)\R\,\bmv_1
+\frac{3\pi}{4}\bmv_1^\T\invdT\K\invdT\R\,\bmt
\nn\\
&
+\frac38\bmt^\T\invdT\left(
1+\frac{5\pi^2}{12}\K\invdT\K\invdT\right)\R\,\bmt ,
\end{align}
which together with (\ref{eq:simpleG}) gives
\begin{equation}
H(T,\bmt)=\frac{9}{16}\bmv_1^\T\invdT\R\,\bmv_1
+\frac{3\pi}{8}\bmv_1^\T\invdT\K\invdT\R\,\bmt
+\frac{\pi^2}{16}\bmt^\T\invdT\K\invdT\K\invdT\R\,\bmt .
\label{eq:Hconcrete}
\end{equation}
In particular, for $\dT^\C$ (\ref{eq:dT^C}) and $\bmt^\C$
(\ref{eq:t^C}),
the corresponding $H$ is given by
\begin{align}
H(T^\C,\bmt^\C)=\frac{9}{8\pi}\,\bmv_1^\T\biggl(
\invK\,\R^\C & - 2\,\K\Bigl(\invK\Bigr)^2\R^\C\K\invK
\nn\\
&
+\K\Bigl(\invK\Bigr)^2\K\invK\,\R^\C \K\Bigl(\invK\Bigr)^2\K
\biggr)\bmv_1 ,
\label{eq:H(T^C,t^C)}
\end{align}
with
\begin{equation}
\R^\C=\biggl(\frac34 + \frac14\K\invK\K\invK\biggr)^{-1} .
\label{eq:R^C}
\end{equation}
This $H(T^\C,\bmt^\C)$ agrees with our previous $H$ given by (4.1) of
\cite{HatMorTer} with the substitution $\bmv_1\to(\pi/3)\K\bmu$ and
$\R^\C\to (4/3)\calR$.
Therefore, the tachyon mass determined from (\ref{eq:PhitcalQBPhit=0})
is different from that determined by (\ref{eq:FockWE}).
This is the phenomenon pointed out in \cite{BCFT} by the BCFT
argument.

\section{Potential height problem: a no-go theorem}

We have seen in secs.\ 2 and 3 that the normalization factor
$\calNc$ for $\Psic$, the tachyon mass $\mt$, and so on differ
depending on whether we consider the equation motion for $\Psic$ and
the wave equation for $\Phit$ on the Fock space or on the sliver
space.
In this section, we shall reexamine the potential height problem
by adopting the sliver space strategy.
The potential height problem is the problem whether the energy density
$\calEc$ of the solution $\Psic$ is equal to the D25-brane tension
$T_{25}$. This problem has been studied in the Fock space strategy in
\cite{HatKaw,HatMor,BCFT,HatMorTer} to obtain an unwelcome result
$\calEc/T_{25}=(\pi^2/3)[16/(27\ln 2)]^3\simeq 2.0558$.
As we shall see below the ratio $\calEc/T_{25}$ can be calculated
rather kinematically without knowing the values of twist anomalies.
Since the the ratio  $\calEc/T_{25}$ is a physical quantity and hence
does not depend on the constant $K$ multiplying the action
(\ref{eq:SV}), we shall put $K=1$ in the rest of this section.
Our analysis in this section and appendix \ref{app:dilatation} also
shows that the value of $\ap\mt^2$ is not important and need not be
equal to $-1$ since it can be varied by a dilatation transformation on
$\Psic$ and $\Phit$.

For the state $\Slv(T,\bmt,p)$ (\ref{eq:S(T,t,p)}) we have the
following formula concerning the center-of-mass momentum dependence:
\begin{align}
&\Slv(T,\bmt,p)\cdot\calQ \Slv(T,\bmt,p')=
\calA\exp\left(-\calV\,p^2\right) (2\pi)^{26}\delta^{26}(p+p') ,
\label{eq:ScalQBS}
\\
&\Slv(T,\bmt,p_1)\cdot\bigl(\Slv(T,\bmt,p_2)*\Slv(T,\bmt,p_3)\bigr)
=\calB\exp\biggl(-\calW\sum_{r=1}^3 p_r^2\biggr)
(2\pi)^{26}\delta^{26}\!\left(p_1+ p_2 + p_3\right) ,
\label{eq:SSS}
\end{align}
where $\calA$, $\calB$,$\calV$ and $\calW$ are constants depending on
$T$ and $\bmt$, but their explicit expressions are unnecessary here.
As the classical solution $\Psic$ and the tachyon wave function
$\Phit$, we adopt those given by (\ref{eq:Psic=NcS}) and
(\ref{eq:Phit=calNtS}), respectively (note that
$\Slv(T)=\Slv(T,\bmt,p=0)$).
Then, the equation motion in the sense of
$\Psic\cdot\left(\calQ\,\Psic + \Psic *\Psic\right)=0$
(\ref{eq:SEOM}) determines $\calNc$ as
\begin{equation}
\calNc=-\frac{\calA}{\calB} .
\label{eq:calNc=-A/B}
\end{equation}
We ignore here the problem found in sec.\ 2.2 that there is no
universal $\calNc$ for which the inner product of the equation of
motion with any sliver space states vanishes.

The tachyon mass $\mt$ and the normalization factor $\calNt$ for
$\Phit$ are determined from the condition
$\Phit(p')\cdot\calQB\Phit(p)\simeq (p^2+\mt^2)
(2\pi)^{26}\delta(p+p')$ near the on-shell $p^2\simeq -\mt^2$.
{}From (\ref{eq:ScalQBS}) and (\ref{eq:SSS}), we have
\begin{align}
\Phit(p')\cdot\calQB\Phit(p)
&=\calNt^2\left(\calA\, e^{-\calV p^2}
+2\,\calNc\calB e^{-2\calW p^2}
\right) (2\pi)^{26}\delta^{26}(p+p') ,
\end{align}
which determines $\mt^2$ and $\calNt$ as follows:
\begin{align}
-\mt^2 &=\frac{\ln 2}{2\calW - \calV} ,
\label{eq:mtbyVW}
\\
\Nt &=\frac{1}{\sqrt{\calA(2\calW-\calV)}}\,e^{-\calV\mt^2/2} .
\label{eq:calNtbyVW}
\end{align}
There is no guarantee that the tachyon mass squared $\mt^2$ of
(\ref{eq:mtbyVW}) is equal to $-1$ despite that we are adopting the
convention of $\ap=1$.
Postponing the remedy for this problem for the moment, let us proceed
to the calculation of the ratio $\calEc/T_{25}$.

The energy density $\calEc$ of the solution $\Psic$ is given by
\begin{equation}
\calEc=\frac{1}{(2\pi)^{26}\delta^{26}(p=0)}\SV[\Psic]
=\frac{1}{(2\pi)^{26}\delta^{26}(0)}\frac16\Psic\cdot\calQ\Psic
=\frac16\calNc^2\calA=\frac{\calA^3}{6\calB^2} .
\label{eq:calEc}
\end{equation}
The D25-brane tension $T_{25}$ is given in terms of
the open string coupling constant $\go$ as
$T_{25}=1/(2\pi^2\go^2\alpha^{\prime 3})$, and
$\go$ is defined as the three-tachyon on-shell coupling:
\begin{equation}
\Phit(p_1)\cdot\bigl(\Phit(p_2)*\Phit(p_3)\bigr)|_{p_r^2=-\mt^2}
=\go\,\delta^{26}\left(p_1+ p_2 + p_3\right) .
\label{eq:Phit^3=go}
\end{equation}
Using (\ref{eq:SSS}) we have
\begin{equation}
\go^2=\left(\calNt^3\calB e^{3\calW \mt^2}\right)^2
=\frac{\calB^2}{8\calA^3(2\calW-\calV)^3} .
\label{eq:go^2}
\end{equation}
{}From (\ref{eq:calEc}) and (\ref{eq:go^2}), we obtain the ratio:
\begin{equation}
\frac{\calEc}{T_{25}}=2\pi^2\go^2\alpha^{\prime 3}\calEc
=\frac{\pi^2}{24(2\calW-\calV)^3} .
\label{eq:ratio}
\end{equation}
However, this is not our final answer for the ratio:
we have to take into account the fact mentioned above that $\ap\mt^2$
(\ref{eq:mtbyVW}) is not necessarily equal to $-1$.
To make $\ap\mt^2$ equal to $-1$, we rescale $\ap$ by
$(2\calW-\calV)/\ln 2$. Since the ratio (\ref{eq:ratio}) is
proportional to $\alpha^{\prime 3}$, this rescaling effects
multiplying (\ref{eq:ratio}) by $\left((2\calW-\calV)/\ln 2\right)^3$.
Namely, the ratio in the mass unit with $\ap\mt^2=-1$ is
\begin{equation}
\left.\frac{\calEc}{T_{25}}\right|_{\ap\mt^2=-1}
=\frac{\pi^2}{24(\ln 2)^3}\simeq 1.2348 .
\label{eq:ratio2}
\end{equation}
One might think that the above argument of rescaling $\ap$ is too
abrupt.
In appendix \ref{app:dilatation}, we derive the same result as
(\ref{eq:ratio2}) by constructing via the dilatation transformation a
new solution $\Psic$ and a new wave function $\Phit$ for which we have
$\ap\mt^2=-1$.

Our conclusion in this section is a negative one:
we can never obtain the desired ratio $\calEc/T_{25}=1$ in
the present framework. We have to find a way to avoid
the application of this no-go theorem.
One possibility is the one mentioned in sec.\ 1 that
the string field $\Psi$ of VSFT is a constrained one with non-linear
representation.
If this is the case, the arguments in
this section do not apply since both the tachyon kinetic term
and the three-tachyon coupling have additional contributions due to
non-linearity.

\section*{Acknowledgments}
We would like to thank I.\ Bars, Y.\ Imamura, I.\ Kishimoto, H.\
Kogetsu, Y.\ Matsuo, K.\ Ohmori, Y.\ Okawa, T.\ Okuda, L.\ Rastelli,
S.\ Teraguchi and B.\ Zwiebach for valuable discussions and comments.
The works of H.\,H.\ was supported in part by a Grant-in-Aid for
Scientific Research from Ministry of Education, Culture, Sports,
Science, and Technology (\#12640264).
S.\,M.\ is supported in part by the JSPS Postdoctoral Fellowships for
Research Abroad (\#472).

\vspace{1.5cm}
\centerline{\Large\bf Appendix}
\appendix

\section{Identities among $M_\alpha$ and $\bmv_\alpha$}
\label{app:Mvidentities}

In this appendix we summarize the identities concerning $M_\alpha$ and
$\bmv_\alpha$. First are the linear relations including the twist
transformation property:
\begin{align}
&C M_0 C=M_0,\quad C M_\pm C= M_\mp,\quad
C M_1 C= -M_1 ,
\\
&C\bmv_0=\bmv_0,\quad C\bmv_\pm =\bmv_\mp,\quad
C\bmv_1= -\bmv_1 ,
\\
&M_0+M_++M_-=1,
\label{eq:M+M+M=1}\\
&\bm{v}_0+\bm{v}_++\bm{v}_-=0 .
\label{eq:v+v+v=0}
\end{align}
Due to (\ref{eq:M+M+M=1}) and (\ref{eq:v+v+v=0}), we can take
$(M_0,M_1)$ and $(\bmv_0,\bmv_1)$ as independent.
Then, the following non-linear identities hold among them:
\begin{align}
&[M_0,M_1]=0,\label{[M,M]}\\
&(1-M_0)(1+3M_0)=M_1^2,\label{MoOfMe}\\
&3(1-M_0)\bmv_0+M_1\bmv_1=0,\label{Mv}\\
&3M_1\bmv_0 +(1+ 3M_0)\bmv_1=0,\label{Mv2}\\
&\frac94\,\bmv_0^2+\frac34\bmv_1^2=2\,V_{00}
=\ln\!\left(\frac{3^3}{2^4}\right) .
\label{vv}
\end{align}

\section{Derivation of (\ref{eq:expanT*T})}
\label{app:derivation}

In this appendix, we present a derivation of the formula
(\ref{eq:expanT*T}).
In the following calculation we strictly keep the ordering of the
matrices and never use the non-linear identities.

First, we shall calculate the action of $(1-T\calM)^{-1}$ on the base
vectors $\cvec{1}{\pm 1}$.
For this purpose,
let us decompose $1-T\calM$ into two parts, $S$ and $A$
(see sec.\ 3.2 of \cite{HatMor}),
\begin{equation}
S=\Pmatrix{1-TM_0&-T(1-M_0)/2\\-T(1-M_0)/2&1-TM_0},\qquad
A=\Pmatrix{0&-TM_1/2\\TM_1/2&0},
\end{equation}
and consider the series expansion with respect to $A$:
\begin{equation}
\frac{1}{1-T\calM}=\frac{1}{S}-\frac{1}{S}A\frac{1}{S}
+\frac{1}{S}A\frac{1}{S}A\frac{1}{S}+\cdots.
\label{Aexpansion}
\end{equation}
As was explained in \cite{HatMor},
higher order terms in this expansion give less singular contributions.
The actions of $1/S$ and $A$ on $\cvec{1}{\pm 1}$ are
\begin{equation}
\frac{1}{S}\cvec{1}{\pm 1}=\cvec{1}{\pm 1}\invS_\pm ,
\qquad
A\cvec{1}{\pm 1}=\mp\cvec{1}{\mp 1}\frac{T M_1}{2} ,
\label{eq:1/SA}
\end{equation}
with
\begin{equation}
\invS_\pm=\left[1-T M_0 -\frac{T(1-M_0)}{2}(1-T M_0)^{-1}
\frac{T(1-M_0)}{2}\right]^{-1}
\left(1\pm \frac{T(1-M_0)}{2}(1-T M_0)^{-1}\right) .
\label{eq:Upm}
\end{equation}
Eq.\ (\ref{eq:1/SA}) for $1/S$ is due to the formula of the inverse
of a $2\times 2$ matrix with matrix components (in our case,
$\calA=\calD$ and $\calB=\calC$):
\begin{equation}
\Pmatrix{\calA& \calB\\ \calC& \calD}^{-1}
=\Pmatrix{
(\calA-\calB\calD^{-1}\calC)^{-1}&(\calC-\calD\calB^{-1}\calA)^{-1}\\
(\calB-\calA\calC^{-1}\calD)^{-1}&(\calD-\calC\calA^{-1}\calB)^{-1}} .
\end{equation}
Using (\ref{eq:1/SA}) in (\ref{Aexpansion}), we get
\begin{equation}
(1-T\calM)^{-1}\cvec{1}{\pm 1}=\left[
\cvec{1}{\pm 1}\invS_\pm
\pm \cvec{1}{\mp 1}\invS_\mp\frac{T M_1}{2}\invS_\pm\right] \D_\pm ,
\label{eq:1/(1-TM)base}
\end{equation}
with $\D_\pm$ defined by
\begin{equation}
\D_\pm =\left(
1+\frac{T M_1}{2}\invS_\mp\frac{T M_1}{2}\invS_\pm\right)^{-1} .
\label{eq:Wpm}
\end{equation}
Then, plugging the decomposition
\begin{equation}
\cvec{M_-}{M_+}=\cvec11\frac{1-M_0}{2}-\cvec{1}{-1}\frac{M_1}{2} ,
\end{equation}
and a similar one for $(M_+,M_-)$ into the definition of $T\star T$,
and using (\ref{eq:1/(1-TM)base}), we obtain an expression of
$T\star T$:
\begin{align}
T\star T = M_0 &+\left(1-M_0 + M_1 \invS_-\frac{TM_1}{2}\right)
\invS_+ \D_+ \frac{T(1-M_0)}{2}
\nn\\
&
+\left((1-M_0)\invS_+\frac{TM_1}{2}- M_1\right)\invS_
-\D_-\frac{TM_1}{2} .
\label{eq:T*T_noapprox}
\end{align}

Up to this point we have made no approximations at all.
To obtain the expansion (\ref{eq:expanT*T}), we use (\ref{eq:expanM0})
and (\ref{eq:T=-1+dT}) in (\ref{eq:Upm}) to get
\begin{equation}
\invS_+ \simeq \frac34\left(1+\frac14 \dT\right),
\qquad
\invS_- \simeq \frac{1}{\dT} .
\label{eq:expanUpm}
\end{equation}
Note that the $O(\K^2)$ term of $M_0$ does not contribute to $\invS_\pm$
to the order given in (\ref{eq:expanUpm}).
Eq.\ (\ref{eq:expanT*T}) is immediately obtained by plugging these
expansions into (\ref{eq:T*T_noapprox}).

The $\star$-product (\ref{eq:T*T}) can naturally be generalized to
$T_1\star T_2$ for two different $T_i$ ($i=1,2$) via the relation
\begin{equation}
\Slv(T_1)*\Slv(T_2)\propto\Slv(T_1\star T_2) ,
\label{eq:S(T1)*S(T2)}
\end{equation}
up to factors and the quantity corresponding to $\hQ$ in
(\ref{eq:S*S}).
Explicitly we have
\begin{equation}
T_1\star T_2 \equiv
M_0+(M_+,M_-)\left[1-\Pmatrix{T_1&0\\0&T_2}
\calM\right]^{-1}\Pmatrix{T_1&0\\0&T_2}\Pmatrix{M_-\\M_+} .
\label{eq:T1*T2}
\end{equation}
Substituting $T_i\simeq -1+\dT_i$ ($i=1,2$) into (\ref{eq:T1*T2}),
we obtain the following formula which is a generalization of
(\ref{eq:expanT*T}):
\begin{align}
T_1\star T_2 \simeq -1 & + \frac14\left(\dT_1+\dT_2\right)
\nn\\
&
+\frac94\left[M_1-\frac13\left(\dT_1-\dT_2\right)\right]
\frac{1}{\dT_1+\dT_2}\left[M_1+\frac13\left(\dT_1-\dT_2\right)
\right] .
\label{eq:expanT1*T2}
\end{align}
Eq.\ (\ref{eq:expanT1*T2}) is derived in quite a similar manner by
considering the decomposition
$1-\sPmatrix{T_1&0\\0&T_2}\calM \\ = S+A$ with
$S$ and $A$ given by
\begin{align}
S&=\Pmatrix{1-(T_1+T_2)M_0/2    & -(T_1M_+ +T_2M_-)/2\\
            -(T_1M_+ +T_2M_-)/2 &1-(T_1+T_2)M_0/2} ,
\\
A&=\Pmatrix{-(T_1-T_2)M_0/2      & -(T_1 M_+ -T_2 M_-)/2 \\
            (T_1 M_+ -T_2 M_-)/2 & (T_1-T_2)M_0/2} .
\end{align}

\section{Rederivation of (\ref{eq:ratio2}) by dilatation transformation}
\label{app:dilatation}

In this appendix, we rederive our result (\ref{eq:ratio2}) for the
ratio $\calEc/T_{25}$ by constructing $\Psic$ and $\Phit$ which give
$\mt^2=-1$. In the following, the space-time dimension is denoted by
$d\,(=26)$, and we adopt the convention $\ap=1$.

The key quantity here is the dilatation transformation $\calD$ on open
string fields:
\begin{equation}
\calD =-\frac{i}{2}\int_0^\pi\! d\sigma\left\{
X^\mu(\sigma),P_\mu(\sigma)\right\}
=-\frac{i}{2}\left\{\hx^\mu,\hp_\mu\right\}
+\frac12\sum_{n\ge 1}\left[(a_n)^2 - (a_n^\dagger)^2\right] ,
\end{equation}
where we have used the mode expansion of the string coordinate
$X^\mu(\sigma)$ and its conjugate
$P_\mu(\sigma)=-i\delta/\delta X^\mu(\sigma)$:
\begin{align}
X^\mu(\sigma)&= \hx^\mu +i\sqrt{2}\sum_{n\ge 1}
\frac{1}{\sqrt{n}}\left(a_n -a_n^\dagger\right)^\mu \cos n\sigma ,
\label{eq:X}\\
P_\mu(\sigma)&=\frac{1}{\pi}\,\hp_\mu
+\frac{1}{\pi\sqrt{2}}\sum_{n\ge 1}\sqrt{n}
\left(a_n +a_n^\dagger\right)^\mu \cos n\sigma .
\label{eq:P}
\end{align}
The operators $\hx^\mu$ and $\hp_\mu$ satisfy
$[\hx^\mu,\hp_\nu]=i\delta^\mu_\nu$ and their actions on the
eigenstates $\ket{p}$ of $\hp$ are
\begin{equation}
\hp\ket{p}=p\ket{p},\qquad
\hx\ket{p}=-i\frac{\p}{\p p}\ket{p} .
\end{equation}
Note that $\calD$ is anti-hermitian, $\calD^\dagger=-\calD$.
The three-string vertex with its matter part given by (\ref{eq:Vmatt})
has the following simple property under the dilatation transformation:
\begin{equation}
\sum_{r=1}^3\calD^{(r)}\ket{V}
=\kappa \ket{V},
\label{eq:DV=kV}
\end{equation}
where $\kappa$ is a constant given by
\begin{equation}
\kappa=\frac{d}{2}-\frac12\sum_{n=1}^\infty V_{nn}^{rr}
=\frac{d}{2}-\frac{1}{12} .
\label{eq:kappa}
\end{equation}
Eq.\ (\ref{eq:DV=kV}) is derived by making integration by parts with
respect to $p_r$ and using the following identities of the
Neumann coefficient matrices which, except (\ref{eq:trV0}), are
equivalent to the non-linear identities given in appendix
\ref{app:Mvidentities}:
\begin{align}
&\sum_{t=1}^3\sum_{k=1}^\infty V_{kn}^{tr} V_{km}^{ts}
=\delta_{nm}\delta^{rs} ,
\\
&\sum_{t=1}^3\sum_{k=1}^\infty V_{kn}^{tr} V_{k0}^{ts}
=V_{n0}^{rs} ,
\\
&\sum_{t=1}^3\sum_{n=1}^\infty V_{n0}^{tr}V_{n0}^{ts}=2 V_{00}^{rs} ,
\\
&\sum_{n=1}^\infty V_{nn}^{rr}=\frac16 ,\quad (r:\mbox{fixed}) .
\label{eq:trV0}
\end{align}
The last identity (\ref{eq:trV0}) has been proved in \cite{Oku2}.

Let us define the dilatation transform of the state $\Slv(T,\bmt,p)$
by
\begin{equation}
\ket{\Slv^\ld(T,\bmt,p)}=e^{\lambda\calD}
\ket{\Slv(T,\bmt,e^\lambda p)} ,
\label{eq:S^l}
\end{equation}
where the momentum of $\ket{\Slv}$ on the RHS is chosen to be
$e^\lambda p$ by taking into account that
$\exp\left(-(i\lambda/2)\{\hx^\mu,\hp_\mu\}\right)\ket{p}
\propto \ket{e^{-\lambda}p}$. The dilatation parameter $\lambda$ will
be fixed later.
Then, corresponding to (\ref{eq:ScalQBS}) and (\ref{eq:SSS}), we have
\begin{align}
&\Slv^\ld(T,\bmt,p)\cdot\calQ\Slv^\ld(T,\bmt,p')=
e^{-d\lambda}\calA\exp\left(-\calV e^{2\lambda}p^2\right)
(2\pi)^{26}\delta^{26}(p+p') ,
\label{eq:ScalQBS^l}
\\[7pt]
&\Slv^\ld(T,\bmt,p_1)\cdot\bigl(
\Slv^\ld(T,\bmt,p_2)*\Slv^\ld(T,\bmt,p_3)\bigr)
\nn\\
&\hspace*{20ex}
=e^{-(d+\kappa)\lambda}\calB\exp\biggl(
-\calW e^{2\lambda}\sum_{r=1}^3 p_r^2\biggr)
(2\pi)^{26}\delta^{26}\Bigl(\sum_r p_r\Bigr) .
\label{eq:SSS^l}
\end{align}
Note that the RHSs of (\ref{eq:ScalQBS^l}) and (\ref{eq:SSS^l}) are
effectively obtained from those of (\ref{eq:ScalQBS}) and
(\ref{eq:SSS}) by the replacements:
\begin{equation}
\calA\to e^{-d\lambda}\calA,\quad
\calB\to e^{-(d+\kappa)\lambda}\calB,\quad
\calV\to e^{2\lambda}\calV,\quad
\calW\to e^{2\lambda}\calW .
\label{eq:repl}
\end{equation}
We consider a new solution $\Psic^\ld$ and the corresponding tachyon
wave function $\Phit^\ld$ in the following form:
\begin{equation}
\Psic^\ld= \calNc^\ld\, S^\ld(T,\bmt,p=0),\quad
\Phit^\ld(p) = \calNt^\ld\, S^\ld(T,\bmt,p) .
\end{equation}
The normalization factor $\calNc^\ld$ determined by
$\Psic^\ld\cdot\bigl(\calQ\,\Psic^\ld + \Psic^\ld *\Psic^\ld\bigr)=0$,
and the tachyon mass squared $(\mt^\ld)^2$ and the normalization
factor $\calNt^\ld$ from the condition
\begin{align}
\Phit(p')\cdot\calQB\Phit(p)
\underset{p^2\sim -(\mt^\ld)^2}{\simeq}
\left(p^2 + \bigl(\mt^\ld\bigr)^2\right)\,
(2\pi)^{26}\delta^{26}(p+p') ,
\end{align}
are given respectively by $\calNc$ (\ref{eq:calNc=-A/B}), $\mt^2$
(\ref{eq:mtbyVW}), and $\calNt$ (\ref{eq:calNtbyVW}) with the
replacements (\ref{eq:repl}).
In particular, the new tachyon mass squared reads
\begin{equation}
-\bigl(\mt^\ld\bigr)^2 =
\frac{\ln 2}{2\calW - \calV}\,e^{-2\lambda} ,
\label{eq:mt^l}
\end{equation}
and we take the following $\lambda$ which realizes $(\mt^\ld)^2=-1$:
\begin{equation}
e^{2\lambda}=\frac{\ln 2}{2\calW - \calV} .
\label{eq:lambda}
\end{equation}
The ratio $\calEc^\ld/T_{25}^\ld$ for the present solution, which is
(\ref{eq:ratio}) with the replacements (\ref{eq:repl}), is given under
the choice (\ref{eq:lambda}) of $\lambda$ by
\begin{equation}
\frac{\calEc^\ld}{T^\ld_{25}}
=\frac{\pi^2}{24(2\calW-\calV)^3}\,e^{-6\lambda}
=\frac{\pi^2}{24(\ln 2)^3} .
\label{eq:ratio^l}
\end{equation}
This is nothing but our previous result (\ref{eq:ratio2}).

\end{document}